%% file: main.tex
\documentclass[aps,pra,superscriptaddress,10pt]{revtex4-2}

\input{preamble.tex}

\begin{document}

\title{Quantum lattice Boltzmann algorithm for heat transfer with phase change}

\author{Christopher L.\ Jawetz}
\affiliation{George W.\ Woodruff School of Mechanical Engineering, Georgia Institute of Technology, Atlanta, GA 30332}

\author{Zhixin Song}
\affiliation{School of Physics, Georgia Institute of Technology, Atlanta, GA 30332}

\author{Spencer H.\ Bryngelson}
\affiliation{George W.\ Woodruff School of Mechanical Engineering, Georgia Institute of Technology, Atlanta, GA 30332}
\affiliation{School of Computational Science \& Engineering, Georgia Institute of Technology, Atlanta, GA 30332}

\author{Alexander Alexeev}\thanks{alexander.alexeev@me.gatech.edu}
\affiliation{George W.\ Woodruff School of Mechanical Engineering, Georgia Institute of Technology, Atlanta, GA 30332}

\begin{abstract}

Heat transfer involving phase change is computationally intensive due to moving phase boundaries, nonlinear computations, and time step restrictions.
This paper presents a quantum lattice Boltzmann method (QLBM) for simulating heat transfer with phase change.
The approach leverages the statistical nature of the lattice Boltzmann method (LBM) while addressing the challenges of nonlinear phase transitions in quantum computing.
The method implements an interface-tracking strategy that partitions the problem into separate solid and liquid domains, enabling the algorithm to handle the discontinuity in the enthalpy-temperature relationship.
We store phase change information in the quantum circuit to avoid frequent information exchange between classical and quantum hardware, a bottleneck in many quantum applications. 
Results from the implementation agree with both classical LBM and analytical solutions, demonstrating QLBM as an effective approach for analyzing thermal systems with phase transitions.
Simulations using 17~lattice nodes with 51~qubits demonstrate root-mean-square (RMS) errors below 0.005 when compared against classical solutions.
The method accurately tracks interface movement during phase transition.
\end{abstract}

\maketitle

\section{Introduction}

Modeling phase transitions in materials presents computational challenges, particularly when tracking the evolving interfaces between solid and liquid states.
The computational demands make such simulations promising candidates for quantum computing approaches, which could offer substantial speedups over classical methods.
Classical computational approaches to phase change problems employ interface-handling strategies with different trade-offs.
Fixed-enthalpy methods~\citep{voller1981} reformulate the energy equation using enthalpy rather than temperature, avoiding explicit interface tracking but often introducing numerical diffusion.
Direct interface-tracking approaches, including front-tracking methods~\citep{UNVERDI199225} and fixed-grid transformations~\citep{voller1990fixed}, maintain interface sharpness but require complex algorithms for topological changes.
Alternative formulations such as level-set methods~\citep{CHEN19978} and phase-field methods~\citep{karma1996phase} offer different balances between accuracy and computational efficiency.
Despite these advances, classical methods face scaling limitations.
This paper develops a quantum lattice Boltzmann method for simulating heat transfer with phase change, focusing on implementing nonlinear phase transitions in a quantum computing context.

The lattice Boltzmann method (LBM), a mesoscale modeling method based on solutions to the Boltzmann equation, offers advantages for quantum implementation~\citep{li2025potential}.
LBM's inherent locality and statistical nature make it well-suited for quantum computing architectures.
Multiphase LBM methods have evolved through several approaches to interface modeling.
\citet{gunstensen1991lattice} developed a multi-component flow method with surface tension derived from a secondary collision step.
\citet{shan1993lattice} used a pseudopotential term to model surface tension between phases with an equation of state in the equilibrium distribution.
Free energy methods by \citet{swift1995lattice} later provided more consistent thermodynamic formulations.
For phase change specifically, the implicit enthalpy method incorporates a phase change term in the collision operator to capture melting effects~\citep{Eshraghi2012}.
This approach avoids costly iterations but still faces computational limits on classical hardware.

In this context, quantum computing has emerged as an alternative to classical approaches for simulating complex physical systems.
Recent advances in quantum hardware from IBM, Microsoft, and Google have improved qubit coherence times and gate fidelity, enabling more complex quantum algorithms~\citep{gupta2024encoding, Majorana, GoogleQuantumAI2025, kikuchi2023realization, ibm2025sqd}.
Some algorithms offer potential exponential speedup over classical methods~\citep{shor1994algorithms, gidney2025factor}.
The Harrow--Hassidim--Lloyd algorithm for solving linear systems exemplifies this potential~\citep{Harrow_2009}.
More advanced quantum linear system algorithms~\citep{morales2024quantum} have been adopted to solve various linear PDEs.
Combined with linearization techniques~\citep{kowalski1991nonlinear}, these algorithms can also solve some nonlinear PDEs~\citep{liu2021efficient, li2025potential, liu2023efficient}.
However, practical implementation on near-term quantum hardware faces limitations from quantum circuit depth.
These solvers also require high-fidelity operations enabled by quantum error correction~\citep{zheng2024}.
These challenges increase when extending the algorithm to nonlinear systems using linearization techniques, which expands the system size and circuit complexity.

Variational quantum algorithms represent another research direction for implementing quantum linear system solvers on near-term hardware~\citep{cerezo2021variational,VQLS}. These algorithms use shallow, parameterized circuits and classical optimization to approximately solve linear systems.
Their shallow depth and robustness against coherent noise make them suitable for noisy intermediate-scale quantum (NISQ) devices.
\citet{liu2022} solved the Laplace equation in one and two dimensions using this method, while \citet{song2025incompressible} solved the incompressible Navier--Stokes equation and tested it on IBM's quantum computer.

Another approach has been through mesoscale modeling techniques such as the lattice Boltzmann method, which are common in classical computing because of their speed and parallelizability.
\citet{Yepez2001} developed the first quantum formulation.
This quantum lattice-gas model mapped each velocity channel to a qubit and implemented streaming through SWAP gates and local BGK-like collisions through unitary transformations.
\citet{Budinski2021} extended this work with the first complete circuit implementation.
They used a linear combination of unitaries approach to implement the non-unitary collision operator and solve the advection--diffusion equation.
Later, they applied the scheme to the stream-function--vorticity formulation of the Navier--Stokes equations~\citep{Budinski2022}.
To address collision nonlinearity, \citet{Itani2022} introduced Carleman linearization that embeds the BGK operator into a larger linear space.
They recently produced a quantum algorithm for lattice Boltzmann implementing both streaming and collision steps as unitary operators~\citep{Itani2024}.
\citet{sanavio2025carleman} explored the relative benefits of Carleman linearization using gate-based and block-encoding techniques.

Recent efforts for QLBM have focused on resource optimization and scalability.
\citet{Kocherla_2024} developed an algorithm that avoids repeated measurements and reinitialization between time steps.
\citet{lee2025multiple} created a two-circuit streamfunction--vorticity algorithm that calculates quantum circuits concurrently and reduces CNOT gate count by 35\% while maintaining $\mathcal{O}(\log N)$ scaling in qubit count.
\citet{Wang2025} introduced a meso-ensemble method with smaller dimensionality than the full cellular automata model while achieving linear collisions.
Other resource-reduction strategies include \citet{Kumar2024}'s method to solve a timestep with a single unitary transformation and \citet{Wawrzyniak2024}'s sparse-matrix encoding, scaled to a $128 \times 128$ lattice on a statevector simulator.

The presented work focuses on whether quantum lattice Boltzmann methods can simulate heat transfer problems with phase change.
The approach introduces a quantum interface-tracking scheme inspired by classical enthalpy approaches to encode the liquid fraction and phase-boundary position as qubit registers.
This scheme allows one to handle the discontinuous enthalpy--temperature mapping at melting.
To our knowledge, this is the first approach to studying phase change problems through QLBM.

This manuscript continues as follows.
\Cref{section:pd} defines the Stefan problem.
\Cref{section:nm} describes the lattice Boltzmann method and its adaptation for phase change.
\Cref{section:qa} describes the quantum lattice Boltzmann method.
\Cref{section:qlbm} presents model additions to incorporate phase change.
\Cref{section:ver} presents the results, verifying the model against analytical and classical LBM solutions.
\Cref{section:conc} provides concluding remarks on the results and their significance.

\section{Problem Description}\label{section:pd}

\begin{figure}[htb]
    \centering
    \includegraphics[]{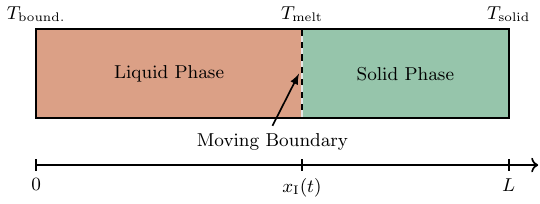}
    \caption{
        Schematic of a 1D phase change problem with a moving interface boundary.
        The domain of length $L$ contains liquid (left) and solid (right) phases separated by a moving boundary at $x_\mathrm{I}(t)$.
        This setup illustrates the Stefan problem for melting processes.
    }
    \label{fig:bar_diagram}
\end{figure}

The analysis considers transient one-dimensional heat transfer with phase change in a finite solid wire at initial temperature $T_\mathrm{solid}$, known as the Stefan problem~\citep{mathModel}.
At time $t=0$, a constant temperature $T_\mathrm{bound}$ is imposed at $x=0$ as illustrated in \cref{fig:bar_diagram}.
The bar melts at temperature $T_\mathrm{melt}$ where $T_\mathrm{solid}<T_\mathrm{melt}<T_\mathrm{bound}$ and creates a moving phase interface at position $x_\mathrm{I}(t)$.
The problem uses an enthalpy $H$ formulation
\begin{equation}
    \frac{1}{c_p}\frac{\partial H}{\partial t}=\alpha \nabla^2 T,
    \label{e:equation}
\end{equation}
where $\nabla^2$ is the spatial Laplacian, $\alpha$ is the heat diffusivity, and $c_p$ is the specific heat capacity at constant pressure~\citep{voller1981}.
The enthalpy $H$ relates to temperature $T$ as
\begin{equation}
    H =
    \begin{cases}
        c_p T, & T < T_\mathrm{melt}, \\
        c_p T + \mathcal{L}_\mathrm{melt}, & T > T_\mathrm{melt},
    \end{cases}
    \label{H-T}
\end{equation}
where $\mathcal{L}_\mathrm{melt}$ is the latent heat of melting.
At $T=T_\mathrm{melt}$, the function becomes discontinuous as enthalpy decouples from temperature.
This decoupling creates a nonlinear relation between $H$ and $T$.
Nonlinear coupling complicates quantum algorithms as $T$ cannot be directly computed from $H$ without quantum measurements and classical computation, which collapses the quantum state.

\citet{rubinstein1971stefan} provides an analytical solution for the classical Stefan problem that tracks the moving phase boundary without a temperature--enthalpy inversion.
The interface position is found as
\begin{equation}
    x_\mathrm{I}(t)=2 \lambda \sqrt{\alpha t},
    \label{eq:interface_position}
\end{equation}
and the interface movement speed $\lambda$ follows from
\begin{equation}
    \lambda \sqrt{\pi} = \frac{\mathrm{St}_\mathrm{liq}}{\exp(\lambda^2) \, \mathrm{erf}(\lambda)} -  \frac{\mathrm{St}_\mathrm{solid}}{\exp(\lambda^2) \, \mathrm{erfc}(\lambda)},
    \label{e:stefan-eqn}
\end{equation}
with $\mathrm{erf}(\cdot)$ and $\mathrm{erfc}(\cdot)$ the error and complementary error functions.
The Stefan numbers $\mathrm{St}$ for liquid and solid phases are
\begin{equation}
    \mathrm{St}_\mathrm{liq}=\frac{c_p(T_\mathrm{bound}-T_\mathrm{melt})}{\mathcal{L}_\mathrm{melt}} \quad \text{and} \quad \mathrm{St}_\mathrm{solid}=\frac{c_p(T_\mathrm{melt}-T_\mathrm{solid})}{\mathcal{L}_\mathrm{melt}}.
\end{equation}
The liquid temperature is
\begin{equation}
    T(x,t)=T_\mathrm{bound}-(T_\mathrm{bound}-T_\mathrm{melt})\frac{\mathrm{erf}({x}/{2\sqrt{\alpha t}})}{\mathrm{erf}(\lambda)}
\end{equation}
and the solid temperature is
\begin{equation}
    T(x,t) = T_\mathrm{melt} - \left( T_\mathrm{melt} - T_\mathrm{solid} \right) \Bigg( \frac{x}{L}
        + \sum_{n=1}^{\infty} \left( \frac{2}{n \pi} \right)
        \sin \left( n \pi \frac{x}{L} \right)
        \exp \left( - \alpha \left( \frac{n \pi}{L} \right)^2 t \right) \Bigg).
    \label{e:temp}
\end{equation}
In \cref{e:temp}, we truncate the infinite sum after 10 terms ($n = 1,2,\dots,10$), which provides sufficiently accurate results for our analysis. 

\section{Computational Model}\label{section:nm}

Equation \cref{e:equation} can be solved using the lattice Boltzmann method (LBM), a mesoscale modeling technique for transport equations~\citep{chen1998lattice}.
Although the governing equations presented above are general for three-dimensional problems, we focus on the one-dimensional case for this study.
LBM models heat transport using fictitious particles moving along a fixed in space lattice. The particles are characterized by a probability density function $f_i(\mathbf{x},t)$.
The algorithm consists of collision and streaming steps.
The collision procedure represents an instant collision of particles arriving from different lattice directions to a lattice node where they collide and relax toward a local equilibrium $f^\mathrm{eq}_i$.
The streaming procedure represents particles moving along lattice links in their respective directions.

The discrete Boltzmann equation describes the evolution of the distribution function $f_i$ as
\begin{equation}
    f_i(\mathbf{x}+\mathbf{e}_i\Delta t,t+\Delta t)=(1-\omega \Delta t)f_i(\mathbf{x},t)+\omega \Delta t  f^\mathrm{eq}_i-w_i \Phi,
\end{equation}
where $\mathbf{x}$ is the lattice position vector, $\Delta t$ is the time step, $f^\mathrm{eq}_i$ is the equilibrium distribution function, $\omega$ is the relaxation coefficient, and $\mathbf{e}_i$ is the lattice velocity vector for direction $i$.
The source term is related to the latent heat of melting as
\begin{equation}
    \Phi=\frac{\dd \eta}{\dd t} \frac{\mathcal{L}_\mathrm{melt}}{c_p} \delta (T-T_\mathrm{melt}).
    \label{phi}
\end{equation}
Here, $\eta$ is the liquid fraction at the lattice node, and $\delta()$ is the Dirac delta function that accounts for latent heat absorption during phase change~\citep{voller1981}.
The implementation uses a time step of $\Delta t = 1$ and lattice spacing of $\Delta x = 1$, and all quantities are dimensionless by the state variables.
The model uses a 1-direction space, 3-velocity (D1Q3) lattice, where the velocity vectors are
$\mathbf{e}_0 = 0$, $\mathbf{e}_1 = -1$, and $\mathbf{e}_2 = +1$,
incorporating rest particles and motion in both directions, creating a balanced lattice structure for stability and isotropy.
The relaxation coefficient $\omega = 2/(6 \alpha + \Delta t)$ is a measure of heat diffusivity~\citep{Eshraghi2012}.

The macroscopic temperature $T$ is calculated from the distribution function as
\begin{equation}
    T(\mathbf{x},t)=\sum_{i=0}^{2}f_i(\mathbf{x},t).
\end{equation}
The equilibrium distribution function is
$f^\mathrm{eq}_i(\mathbf{x},t)= w_iT(\mathbf{x},t)$, with weights $\mathbf{w}=[1/6, ~2/3, ~1/6]^\top$.
The source term $\Phi$ is zero everywhere except at $T=T_\mathrm{melt}$.
At $T=T_\mathrm{melt}$, we calculate $\Phi$ by computing $f_i(\mathbf{x},t+\Delta t)$ without $\Phi$, then setting
\begin{equation}
    \Phi=\sum_{i=0}^{2} f_i(\mathbf{x},t+\Delta t)-T_\mathrm{melt}.
\end{equation}
This relationship enables the liquid fraction change to be evaluated using \cref{phi}.
We scale $f_i(\mathbf{x},t+\Delta t)$ so that $\sum_{i=0}^{2} f_i(\mathbf{x},t+\Delta t)=T_\mathrm{melt}$, preserving energy balance at the interface.
We store the liquid fraction with the temperature, so the enthalpy $H$ is
\begin{equation}
    H = c_p T + \mathcal{L}_\mathrm{melt} \eta,
\end{equation}
corresponding to the enthalpy relation in \cref{H-T}.
We track interface position $x_\mathrm{I}$ with sub-grid precision as
\begin{equation}
    x_\mathrm{I}= x+(\eta-0.5)\Delta x,
\end{equation}
within the melting node at position $x$.

The one-dimensional domain is $x \in [0,L]$ with constant temperature boundary conditions $T(x=0,t)=T_\mathrm{bound}$ and $T(x=L,t)=T_\mathrm{solid}$.
For lattice directions entering from outside the domain, we compute the unknown distribution function values by preserving the boundary temperature as
\begin{equation}
    f_j = T - \sum_{i\neq j} f_i,
\end{equation}
where $f_j$ is the distribution function for lattice direction $j$.

\section{Quantum Algorithm}\label{section:qa}

\begin{figure}[htb]
    \input{figs/QLBM_circuit}
    \caption{
        Quantum circuit for a single time step of the QLBM algorithm between lattice sites $k$ and $k+1$ under the D1Q3 scheme.
        The overall circuit operates on $3N$ qubits initially in the $\ket{0}$ state, where $N$ is the total number of lattice sites.
        Each time step includes five stages: (1) Encoding: initialization of qubit states to represent particle distribution functions using $R_y$ gates, (2) Collision: unitary collision operators $U_c$ modeling particle interactions inside each lattice node $k$, (3) Streaming: particle propagation using SWAP operations between adjacent lattice sites, (4) Measurement: measure the quantum states to retrieve $f_i$; and (5) Reinitialization: prepare the initial quantum state for the next time step based on distribution from previous time step and enforcing boundary conditions. 
    }
    \label{fig:original_circuit}
\end{figure}
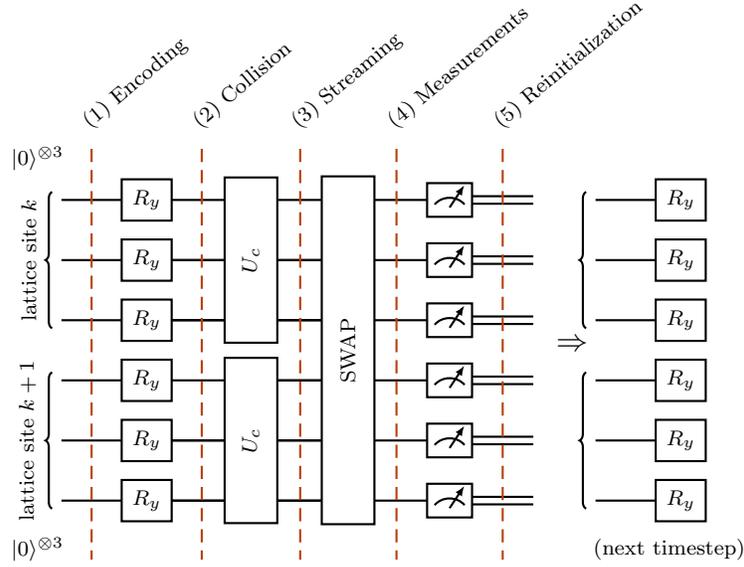

\citet{Yepez2001} introduced the first quantum lattice Boltzmann method (QLBM).
A measurement-based algorithm encodes each particle at each lattice node as a qubit.
The approach is shown in \cref{fig:original_circuit}, where each quantum gate represents a specific operation in the QLBM algorithm.
The circuit involves $R_y$ gates, which perform Y-rotations for encoding particle distribution functions into quantum states using angle $\theta_{k,i}=2\cos^{-1}(\sqrt{1-f_i(x_k,t)})$.
The unitary collision operator $U_c$ models particle interactions during the collision step.
SWAP gates implement the streaming operation for particle propagation between neighboring lattice sites.

The algorithm begins with encoding.
We encode initial temperature distribution values into registers, with the interface node encoded separately in each register.
Each qubit encodes the distribution value $f_i$ as
\begin{equation}
    \ket{q_i(x_k,t)}= \sqrt{1-f_i(x_k,t)} \ket{0} + \sqrt{f_i(x_k,t)} \ket{1}, \quad k=0,1,2,\dots,N-1, \quad i=0,1,2,
    \label{e:encoding}
\end{equation}
where $N$ is the total number of lattice sites, $k$ is the lattice site index, and $i$ is the velocity direction index.
We implement this using an $R_y(\theta_{k,i})$ gate with angle $\theta_{k,i}=2\cos^{-1}(\sqrt{1-f_i(x_k,t)})$ for the ket vector $\ket{q_i(x_k,t)}$.
The state at lattice site $k$ is
\begin{equation}
    \ket{\psi(x_k,t)}= \ket{q_{0}(x_k,t)} \otimes \ket{q_{1}(x_k,t)} \otimes \ket{q_{2}(x_k,t)}.
    \label{e:lattice-state}
\end{equation}
Hence, the overall quantum state is a product state of each lattice site state as
\begin{equation}
    \ket{\psi(x_0,x_1,\cdots,x_{N-1},t)}= \bigotimes_{k=0}^{N-1} \ket{\psi(x_k,t)}.
    \label{e:system-state}
\end{equation}
The QLBM collision operation $U_c$ applies a unitary transformation to each lattice node $\ket{\psi(x_k,t)}$, relaxing the distribution toward equilibrium. 
The operator for D1Q3 is an expansion of the collision operator described in \citet{Yepez2002} for the D1Q2 scheme. 
To conserve energy within the node during the collision, this operator disjointly entangles the states of $\mathcal{S}_1 =\{\ket{001}, \ket{010}$, $\ket{100}\}$ and $\mathcal{S}_2=\{\ket{011}, \ket{101}$, $\ket{110}\}$. 
Thus, the $8 \times 8$ matrix $U_c$ is described via two SU(3) unitary group block matrices, which represent the disjoint mixing processes, along with unitary values for $\ket{000}$ and $\ket{111}$. 
Particle-hole symmetry is enforced with a global NOT operation on the second set of states $\mathcal{S}_2$ using a permutation matrix
\begin{equation}
    P = 
    \begin{bmatrix}
    0 & 0 & 1 \\
    0 & 1 & 0 \\
    1 & 0 & 0
    \end{bmatrix}.
\end{equation}
Then, we can write the collision operator $U_c$ as the direct sum of each component
\begin{equation}
    U_c = 1 \oplus V \oplus P V P^\dagger \oplus 1.
\end{equation}
Imposing isotropy on the link directions limits $V$ to linear combinations of the identity matrix $I_{3\times3}$ and the all-ones-matrix $J_{3\times3}$ (normalized to satisfy unit trace $\text{Tr}[J]=1$) as
\begin{equation}
    V = \alpha J + \beta (I-J).
\end{equation}
We require $|\alpha|=|\beta|=1$ to preserve unitarity, resulting in a free parameter $\alpha=1,\beta=\exp(i\theta)$.
We choose $\theta=2\pi/3$ to maximize the mixing between the three quantum states, representing $\omega=1$ in standard LBM heat transfer models.
After reframing the result in the canonical basis, we obtain 
\begin{equation}
    U_c=\frac{1}{\sqrt{3}}
    \begin{bmatrix}
    \sqrt{3} & 0 & 0 & 0 & 0 & 0 & 0 & 0 \\
    0 & i & \exp{(-i\pi/6)} & \exp{(-i\pi/6)} & 0 & 0 & 0 & 0 \\
    0 & \exp{(-i\pi/6)} & i & \exp{(-i\pi/6)} & 0 & 0 & 0 & 0 \\
    0 & 0 & 0 & 0 & \exp{(-i\pi/6)} & \exp{(-i\pi/6)} & i & 0 \\
    0 & \exp{(-i\pi/6)} & \exp{(-i\pi/6)} & i & 0 & 0 & 0 & 0 \\
    0 & 0 & 0 & 0 & \exp{(-i\pi/6)} & i & \exp{(-i\pi/6)} & 0 \\    
    0 & 0 & 0 & 0 & i & \exp{(-i\pi/6)} & \exp{(-i\pi/6)} & 0 \\
    0 & 0 & 0 & 0 & 0 & 0 & 0 & \sqrt{3}
    \end{bmatrix}.
\end{equation}

The streaming operation uses the established quantum walk algorithm, permuting qubits via SWAP gates~\citep{Budinski2021}.
Last, we reset the qubit phases to restore the post-initialization state.
The results of \cref{section:ver} show that one can advance multiple time steps without reinitialization.
By avoiding reinitialization, one also avoids repeated state preparation and readout solution states, which can introduce noise that accumulates with each measurement--reinitialization cycle.

\section{QLBM with Phase Change}\label{section:qlbm}

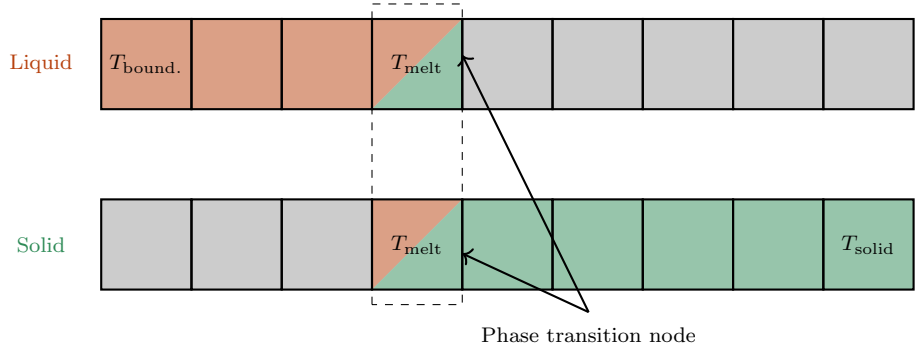
\begin{figure}[ht]
    \centering
    \input{figs/table_figure.tikz}
    \caption{
        The discretization scheme for the phase change problem shows grids for two phases (brown, liquid; green, solid).
        The phase boundary is indicated in the dashed box.
        The boundary conditions are $T_\mathrm{bound}$ and $T_\mathrm{solid}$ for the left and right boundaries.
        The grayed-out nodes are for illustrative purposes, as each phase only exists in one of the two grids or the phase change node, and so are not used in the calculations.
    }
    \label{fig:table}
\end{figure}

To incorporate phase change, we separate the system into two registers at the melting node, each containing a copy of the temperature field as shown in \cref{fig:table}.
The system wavefunction for the liquid region is
\begin{equation}
    \ket{\Psi(x_0,x_1,\dots,x_{N_\mathrm{liq}-1},t)}=\bigotimes_{k=0}^{N_\mathrm{liq}-1}\ket{\psi(x_k,t)}
\end{equation}
and for the solid region is
\begin{equation}
    \ket{\Psi(x_{N_\mathrm{liq}-1},x_{N_\mathrm{liq}},\dots,x_{N-1},t)} = \bigotimes_{k=N_\mathrm{liq}-1}^{N-1}\ket{\psi(x_k,t)}
\end{equation}
for a period with $N_\mathrm{liq}$ liquid sites and $N-N_\mathrm{liq}$ solid sites.

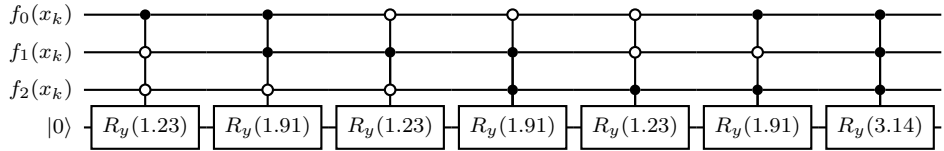
\begin{figure}[htb]
    \input{figs/mcry_circuit_figure}
    \caption{
        MCRY circuit evaluates possible measurement combinations at each lattice node, shown for the distributions $f_i$, by rotating the ancilla qubit.
        Rotation angles are determined by the sum $S$ of control values.
    }
    \label{fig:mcry-circuit}
\end{figure}

We implement a multi-controlled rotation (MCRY) gate sequence at the liquid--solid interface to measure post-streaming temperature within the circuit.
This circuit, shown in \cref{fig:mcry-circuit}, systematically evaluates all possible measurement combinations at each node, applying specific rotation angles to encode temperature information in an ancilla qubit's probability amplitude.
The rotation angle $\theta=2\sin^{-1}(\sqrt{S/3})$ ensures the ancilla qubit yields 1 with probability $S/3$, where $S$ represents the sum of control bit values.
This procedure produces a single rotation each timestep with an amplitude corresponding to the measured value.
The mapping creates a linear correspondence between measured probability and temperature $T$.

The circuit rotation angles correspond to possible combinations of the three input qubits representing discrete temperature distributions.
The corresponding numerical values ($1.23$, $1.91$, and so on) are computed from the inverse sine function applied to the square root of the normalized sum of control values, scaled to match the expected temperature range at the phase interface.
We scale the probability amplitude by $\mathcal{L}_\mathrm{melt} ((S/3)-T_\mathrm{melt})/c_p$ to compute the liquid fraction change, which is applied after measurement.

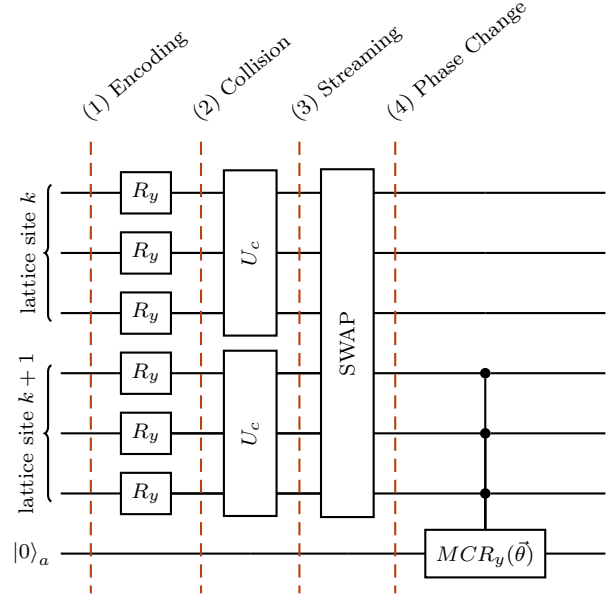
\begin{figure}[ht]
    \centering
    \input{figs/full_circuit_figure}
    \caption{
        QLBM circuit, using the MCRY gate to store phase change data from the boundary lattice site.
        Circuit symbols: $R_y$ are Y-rotation gates; $U_c$ is the unitary collision operator; $C$ denotes control qubits; and $MCR_y(\vec{\theta})$ is the multi-controlled rotation gate for phase change detection.
    }
    \label{fig:full-circuit}
\end{figure}

The MCRY circuit stores information without disrupting the superposition of other qubits, as shown in \cref{fig:full-circuit}.
This QLBM approach is an improvement over traditional QLBM methods, separating the nonlinear contribution without collapsing the quantum state.
The measured phase change information propagates through later time steps via classical tracking of the liquid fraction.
This hybrid quantum--classical approach enables efficient phase change simulation while maintaining the quantum coherence of the primary computation.
The system stores phase change history in the ancilla, enabling accurate modeling of the liquid-solid interface evolution.

The implementation requires a different approach to the boundary conditions.
Each time step produces one unknown lattice direction at fixed boundaries, traditionally calculated using known boundary values.
This calculation cannot rely on direct boundary-value observations without intermediate measurements.

\begin{figure}[ht]
        \includegraphics[]{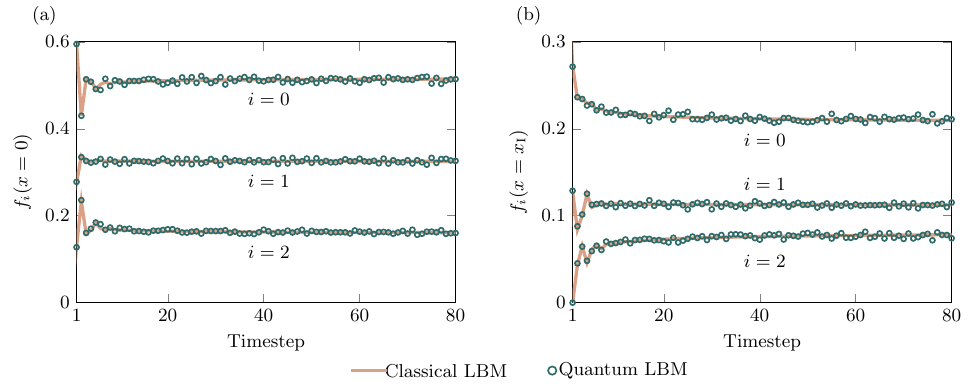}
        \caption{Time evolution of LBM distribution function values at (a) the left boundary and (b) the phase interface.
        Distributions $f_i$ along different lattice directions $i$ are shown as labeled.
        }
        \label{fig:bcs}
\end{figure}

In regions undergoing phase change, the melting node exhibits quasi-static behavior, enabling previous boundary condition values to be used.
We verify this approach by comparing classical and quantum solutions (\cref{fig:bcs}), showing good agreement.
In \cref{fig:bcs}, the value of the distribution functions $f_i$ is shown at the (a) left boundary and (b) the node undergoing melting; the time steps here denote the time after phase change moves into the labeled node (grid cell).
With this, we see that after about 5~time~steps we no longer need to compute boundary condition values as that distribution $f_i$ reaches a steady state.
As the liquid fraction approaches 1, the system reverts to the original algorithm until distribution function values reach an approximate steady state. 
Still, the phase interface moves through the domain as the simulation progresses ($t \in [0,80]$, here).
Thus, this test provides a faithful physical benchmark for algorithm performance evaluation.

We use linear extrapolation of the liquid fraction to predict complete melting at a lattice node, which behaves linearly because of constant boundary conditions.
When the melting node changes, the system reverts to per-time-step measurements until reaching an updated equilibrium state.
The transition period duration varies with thermal conductivity $\alpha$ but remains short for typical metal values.

\section{Results}\label{section:ver}

\begin{figure}[ht]
    \includegraphics[]{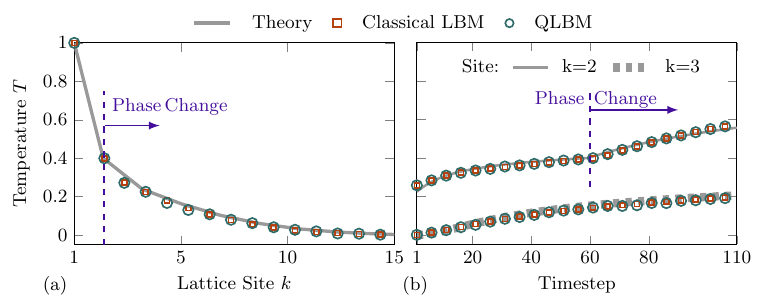}
    \caption{
    Comparison between the analytic solution, classical LBM, and quantum LBM (this work) methods.
    (a) Temperature distribution after 110 time steps, equivalent to $t=110$.
    (b) Temperature time evolution at lattice sites $k = 2$ and $k = 3$. 
    The melting node is initiated at lattice site $k = 1$ and moves from left to right through the simulation until time~step~110.
    The vertical dashed lines indicate the interface location as predicted by theory.
    }
    \label{fig:temp_evolution}
\end{figure}

\Cref{fig:temp_evolution} compares the analytic solution \cref{e:stefan-eqn}, classical LBM solution, and QLBM solution for one-dimensional heat transfer along a bar with $T_\mathrm{melt}=0.4 \, T_\mathrm{bound}$ and $\mathcal{L}_\mathrm{melt}/c_p=10$.
We chose parameters to examine phase change behavior in a moderate temperature regime, with the melting point positioned away from boundaries.
The simulation uses 17~lattice nodes with 51~qubits, the required balance between spatial resolution and the number of qubits.
We perform QLBM simulations via the Qiskit~Aer MPS simulator~\citep{qiskit2024} for 110~time steps ($t \in [0,110]$). 

We find close agreement among the three solutions in terms of temperature distribution shown in \cref{fig:temp_evolution}~(a) and temperature evolution shown in \cref{fig:temp_evolution}~(b), confirming that QLBM can properly solve this problem and represent the interface dynamics.
The results show good agreement between all solutions for the temperature $T(x)$.

The temperature evolution in \cref{fig:temp_evolution}~(b) shows that the simulations using the quantum algorithm represent the phase change dynamics with small errors.
The quantum solution tracks the classical result throughout the phase transition period, representing the initiation and progression of the phase change.
This agreement demonstrates an effective approach to representing nonlinear behavior in a quantum computing context.
Furthermore, the interface position predicted by the quantum solution closely follows the analytical solution $x_\mathrm{I}(t)$ given in \cref{eq:interface_position}.

\begin{figure}[ht]
    \centering
    \includegraphics[]{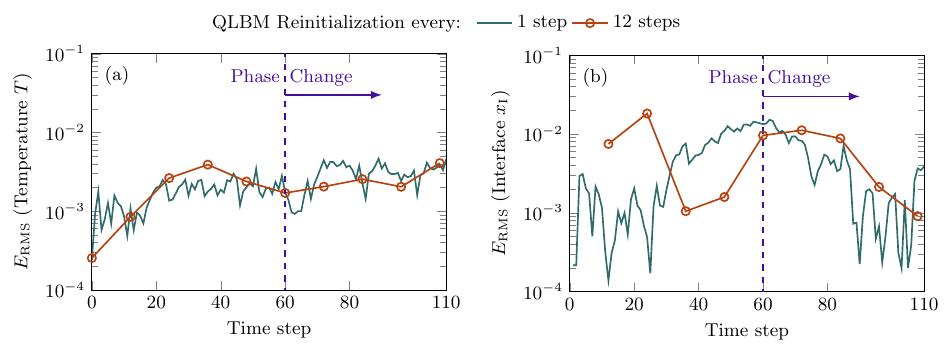}
    \caption{
    Error results comparing quantum and classical LBM solutions and reinitialization schemes.
    The RMS error follows \cref{e:rms} for (a) temperature $T$ and (b) interface position $x_\mathrm{I}$.
    Phase change occurs at step~60, though the errors do not increase.
    }
    \label{fig:errors}
\end{figure}

Error analysis of the quantum solutions is shown in \cref{fig:errors}, key characteristics of the implementation's performance.
We evaluate the error between the quantum circuit and classical LBM solutions.
The absolute root mean square (RMS) error for a quantity $q$ is defined as
\begin{equation}
    E_{\mathrm{RMS}}=\sqrt{\frac{1}{M}\sum_{k=1}^{M}\left(q_{\text{classical}}^{(k)}-q_{\mathrm{quantum}}^{(k)}\right)^2},
    \label{e:rms}
\end{equation}
where $k$ is the lattice site index and $M=17$ is the total number of lattice sites.

The errors for both quantities of interest, temperature $T$ and interface position $x_\mathrm{I}$, remain below $0.005$ throughout the simulation.
The relative and absolute RMS errors are similar in scale, so we do not show both here.
\Cref{fig:errors} also shows that the interface location $x_\mathrm{I}$ error has comparable accuracy while reducing the overhead of quantum measurements when we postpone the reinitialization to every 12 time steps.

\section{Conclusions}\label{section:conc}

This paper presents a quantum lattice Boltzmann method (QLBM) for heat transfer with phase change.
The method demonstrates how quantum algorithms can handle phase change nonlinearities and reduce the computational burden of repeated reinitialization.
Further, using a variable source term via an enthalpy-based formulation extends the method to represent phase change dynamics.

Results show that the presented QLBM technique accurately represents heat transfer with phase change, as verified by comparisons to classical LBM simulations and analytical solutions.
Using the D1Q3 lattice, quantum simulations replicate temperature distribution and liquid fraction evolution during phase transition.
As the phase change and boundary conditions are captured within the circuit, we hope to expand this work to proceed for several time steps without reinitialization. 
Reducing reinitialization count increases computational efficiency at a cost linear in the additional qubits.
The differences between quantum, classical, and analytical solutions remain small across all tests.

Although the QLBM algorithm was originally designed for fault-tolerant quantum computers that can run millions of high-fidelity quantum operations~\citep{preskill2025beyond}, some recent works show that heavy optimization on the algorithm and circuit design can realize simple QLBM runs on current quantum hardware~\citep{tiwari2025algorithmic}.
Adopting those recent developments and the idea of minimizing frequent measurements would further reduce the computational overhead of our QLBM algorithm, making the method more practical for early fault-tolerant quantum computers.
Quantum computing's inherent parallelism opens possibilities for solving complex PDE-based problems more efficiently than classical methods alone.
In this case, we show the technique is well-suited to heat transfer problems with phase transition.
The method naturally extends to two- and three-dimensional problems by using higher-dimensional lattice structures (for example, D2Q9 or D3Q19), although the effects of fluid convection may need to be incorporated into the solution of higher dimensional problems.
It can apply the same interface-tracking strategy across multiple spatial dimensions, with the circuit complexity scaling logarithmically with the number of lattice sites.

This work demonstrates QLBM's suitability for solving complex heat transfer problems with phase change, which are relevant to diverse engineering applications, including thermal management, energy storage, and additive manufacturing.
Addressing nonlinearity and reducing computational overhead advances quantum computing application to simulating physical systems, extending beyond previously explored systems.
These results suggest QLBM extends to a broader range of engineering applications, potentially offering speedups compared to classical methods.

\section*{Acknowledgments}

This work was supported by the CRNCH Fellowship from the Georgia Institute of Technology and the National Science Foundation via grant no.\ CBET~2217647.

\section*{Data availability}
The code for this paper is available at: \url{https://github.com/comp-physics/Quantum_Heat_LBM}.

\bibliography{quantum_bib}

\end{document}

%% file: preamble.tex
\usepackage[utf8]{inputenc}

\usepackage{graphicx}
\usepackage{amsmath}
\usepackage{color}
\usepackage{soul}
\usepackage{booktabs}
\usepackage{amssymb}
\usepackage{empheq}
\usepackage{etoolbox}
\usepackage{stmaryrd}
\usepackage{hyperref}
\usepackage{braket}
\usepackage{bbm}

\usepackage{tikz}
\usepackage{pgfplotstable}
\usepackage{pgfplots}
\usepgfplotslibrary{groupplots}
\usepgfplotslibrary{fillbetween} 
\usetikzlibrary{arrows}
\usetikzlibrary{shapes}
\usetikzlibrary{decorations.text}
\usetikzlibrary{quantikz}
\usetikzlibrary{arrows.meta}
\usetikzlibrary{quotes}
\usetikzlibrary{calc}
\usetikzlibrary{fadings}
\usetikzlibrary{decorations.pathreplacing}
\usetikzlibrary{patterns}
\usetikzlibrary{3d}
\usetikzlibrary{intersections}
\usetikzlibrary{bending}
\usetikzlibrary{cd}
\usetikzlibrary{fit}
\usetikzlibrary{decorations.markings}
\usetikzlibrary{decorations}
\usetikzlibrary{backgrounds}
\usetikzlibrary{positioning}

\usepackage[]{geometry}
\usepackage[]{siunitx}
\definecolor{lightblue}{rgb}{0.63, 0.74, 0.78}
\definecolor{seagreen}{rgb}{0.18, 0.42, 0.41}
\definecolor{orange}{rgb}{0.85, 0.55, 0.13}
\definecolor{silver}{rgb}{0.69, 0.67, 0.66}
\definecolor{rust}{rgb}{0.72, 0.26, 0.06}
\definecolor{purp}{RGB}{68, 14, 156}

\colorlet{lightrust}{rust!50!white}
\colorlet{lightorange}{orange!25!white}
\colorlet{lightlightblue}{lightblue}
\colorlet{lightsilver}{silver!30!white}
\colorlet{darkorange}{orange!75!black}
\colorlet{darksilver}{silver!65!black}
\colorlet{darklightblue}{lightblue!65!black}
\colorlet{darkrust}{rust!85!black}
\colorlet{darkseagreen}{seagreen!85!black}

\usepackage[final,nopatch=footnote]{microtype}

\usepackage{hyperref}
\hypersetup{
  colorlinks=true,
  linkcolor=darkrust,
  citecolor=seagreen,
  urlcolor=darkrust,
  pdfauthor=author,
}

\usepackage[nameinlink]{cleveref}
\crefname{equation}{}{}

\usepackage[parfill]{parskip}

\pgfplotsset{
    compat=1.18,
    standard/.style={
        scale only axis,
        width=0.5\textwidth,
        enlarge x limits=0.05,
        enlarge y limits=0.05,
        max space between ticks=40,
    },
    every axis/.append style={font=\footnotesize},
    every legend/.append style={font=\footnotesize},
    every node/.append style={font=\footnotesize},	
}

\tikzset{
    every node/.append style={font=\footnotesize},	
}

\usepackage{setspace}
\newcommand{\dd}{\mathrm{d}}

%% file: figs/QLBM_circuit.tex
\begin{tikzpicture}[very thick]
\node at (0,0) {
    \begin{quantikz}[row sep={0.8cm,between origins}, column sep=0.8cm, slice style={draw=rust}, slice label
    style={inner sep=0.5pt,anchor=south west,rotate=40}]
    & \lstick[wires=3,label style={rotate=90,anchor=south}]{lattice site $k$}\slice[style={draw=rust}, label style={rotate=40,anchor=south west}]{(1) Encoding} & \gate{R_y}\slice[style={draw=rust}, label style={rotate=40,anchor=south west}]{(2) Collision} & \gate[3,style={minimum width=0.7cm, inner ysep=-0.01cm}]{\rotatebox{90}{$U_c$}}\slice[style={draw=rust}, label style={rotate=40,anchor=south west}]{(3) Streaming} & \gate[6,style={minimum width=0.7cm, inner ysep=-0.26cm}]{\rotatebox{90}{\text{SWAP}}}\slice[style={draw=rust}, label style={rotate=40,anchor=south west}]{(4) Measurements } & \meter{} \slice[style={draw=rust}, label style={rotate=40,anchor=south west}]{(5) Reinitialization}  & \cw \\
    &  & \gate{R_y} & \qw & \qw & \meter{}  & \cw \\
    &  & \gate{R_y} & \qw & \qw & \meter{}  & \cw \\
    & \lstick[wires=3,label style={rotate=90,anchor=south}]{lattice site $k+1$}  & \gate{R_y} & \gate[3,style={minimum width=0.7cm, inner ysep=-0.01cm}]{\rotatebox{90}{$U_c$}} & \qw & \meter{} & \cw \\
    &  & \gate{R_y} & \qw & \qw & \meter{}  & \cw \\
    &  & \gate{R_y} & \qw & \qw & \meter{}  & \cw 
    \end{quantikz}
    \hspace{-1.5cm}
    {\large $\Rightarrow$}
    \hspace{-1.0cm}
    \begin{quantikz}[row sep={0.8cm,between origins}, column sep=0.8cm]
    &\lstick[wires=3]{} & \gate{R_y} \\
    & & \gate{R_y} \\
    & & \gate{R_y} \\
    &\lstick[wires=3]{} & \gate{R_y} \\
    & & \gate{R_y} \\
    & & \gate{R_y} 
    \end{quantikz}
    };
\node at (-4.2,1.5) {$|0\rangle^{\otimes 3}$};
\node at (-4.2,-3.7) {$|0\rangle^{\otimes 3}$};
\node at (4.2,-3.7) {(next timestep)};
\end{tikzpicture} 

%% file: figs/table_figure.tikz
\begin{tikzpicture}

\def\cell{1.2}

\definecolor{seagreen}{RGB}{46,139,87}
\definecolor{rust}{RGB}{183,65,14}

\foreach \row in {0,2} {
  \foreach \i in {0,...,8} {
    \path[fill=gray!40] (\i*\cell, \row*\cell) rectangle ++(\cell,\cell);
  }
}

\foreach \i in {0,1,2} {
  \path[fill=rust!50] (\i*\cell, 2*\cell) rectangle ++(\cell,\cell);
}

\foreach \i in {4,5,6,7,8} {
  \path[fill=seagreen!50] (\i*\cell, 0*\cell) rectangle ++(\cell,\cell);
}

\fill[rust!50] (3*\cell, 2*\cell) -- (4*\cell, 3*\cell) -- (3*\cell, 3*\cell) -- cycle;
\fill[seagreen!50] (4*\cell, 3*\cell) -- (4*\cell, 2*\cell) -- (3*\cell, 2*\cell) -- cycle;
\fill[rust!50] (3*\cell, 0*\cell) -- (4*\cell, 1*\cell) -- (3*\cell, 1*\cell) -- cycle;
\fill[seagreen!50] (4*\cell, 1*\cell) -- (4*\cell, 0*\cell) -- (3*\cell, 0*\cell) -- cycle;

\foreach \row in {0,2} {
  \foreach \i in {0,...,8} {
    \draw[thick] (\i*\cell, \row*\cell) rectangle ++(\cell,\cell);
  }
}

\draw[dashed] (3*\cell, -0.2) rectangle (4*\cell, 3.8);

\node at (-0.8, 2.5*\cell) {\textcolor{rust}{Liquid}};
\node at (-0.8, 0.5*\cell) {\textcolor{seagreen}{Solid}};

\node at (0.5*\cell, 2.5*\cell) {$T_\mathrm{bound.}$};
\node at (8.5*\cell, 0.5*\cell) {$T_\mathrm{solid}$};
\node at (3.5*\cell, 2.5*\cell) {$T_\mathrm{melt}$};
\node at (3.5*\cell, 0.5*\cell) {$T_\mathrm{melt}$};

\draw[->, thick] (5.4*\cell, -0.3) -- (4.0*\cell, 0.4*\cell);
\draw[->, thick] (5.4*\cell, -0.3) -- (4.0*\cell, 2.6*\cell);

\node at (5.4*\cell, -0.6) {Phase transition node};

\end{tikzpicture}

%% file: figs/mcry_circuit_figure.tex
\begin{quantikz}[row sep={0.5cm,between origins}, column sep=0.1cm]
    & \lstick{$f_{0}(x_k)$} & \ctrl{1} & \qw & \ctrl{2} & \qw & \octrl{2} & \qw & \octrl{3} & \qw & \octrl{1} & \qw & \ctrl{1} & \qw & \ctrl{3} & \qw \\
    & \lstick{$f_{1}(x_k)$} & \octrl{1} & \qw & \control{} & \qw & \ctrl{1} & \qw & \ctrl{2} & \qw & \octrl{2} & \qw & \octrl{2} & \qw & \ctrl{2} & \qw \\
    & \lstick{$f_{2}(x_k)$} & \octrl{1} & \qw & \octrl{1} & \qw & \octrl{1} & \qw & \ctrl{1} & \qw & \ctrl{1} & \qw & \ctrl{1} & \qw & \ctrl{1} & \qw \\
    & \lstick{$\ket{0}$} & \gate{R_y(1.23)} & \qw & \gate{R_y(1.91)} & \qw & \gate{R_y(1.23)} & \qw & \gate{R_y(1.91)} & \qw & \gate{R_y(1.23)} & \qw & \gate{R_y(1.91)} & \qw & \gate{R_y(3.14)} & \qw
\end{quantikz}

%% file: figs/full_circuit_figure.tex
\begin{tikzpicture}[very thick]
\node at (0,0) {
    \begin{quantikz}[row sep={0.8cm,between origins}, column sep=0.8cm, slice style={draw=rust}, slice label
    style={inner sep=0.5pt,anchor=south west,rotate=40}]
    & \lstick[wires=3,label style={rotate=90,anchor=south}]{lattice site $k$} \slice[style={draw=rust}, label style={rotate=40,anchor=south west}]{(1) Encoding} & \gate{R_y}\slice[style={draw=rust}, label style={rotate=40,anchor=south west}]{(2) Collision} & \gate[3,style={minimum width=0.7cm, inner ysep=-0.01cm}]{\rotatebox{90}{$U_c$}}\slice[style={draw=rust}, label style={rotate=40,anchor=south west}]{(3) Streaming} & \gate[6,style={minimum width=0.7cm, inner ysep=-0.26cm}]{\rotatebox{90}{\text{SWAP}}} \slice[style={draw=rust}, label style={rotate=40,anchor=south west}]{(4) Phase Change} & \qw & \qw  \\
    &  & \gate{R_y} & \qw  & \qw & \qw & \qw \\
    &  & \gate{R_y} & \qw  & \qw & \qw & \qw \\
    &\lstick[wires=3,label style={rotate=90,anchor=south}]{lattice site $k+1$}  & \gate{R_y} & \gate[3,style={minimum width=0.7cm, inner ysep=-0.01cm}]{\rotatebox{90}{$U_c$}} & \qw  & \ctrl{3} & \qw \\
    &  & \gate{R_y} & \qw & \qw & \ctrl{2} & \qw \\
    &  & \gate{R_y} & \qw & \qw & \ctrl{1} & \qw \\
    & \lstick{$\ket{0}_a$} & \qw  & \qw & \qw & \gate{MCR_y(\vec{\theta})} & \qw
    \end{quantikz}
    };
\end{tikzpicture} 